\documentclass[11pt]{article}
\usepackage[left=2.5cm,right=2.5cm,top=2.5cm,bottom=2.5cm,a4paper]{geometry} 
\usepackage{graphicx}
\usepackage{amsmath,amssymb}
\usepackage{physics}
\usepackage[affil-it]{authblk}

\newcommand{\vevx}[1]{\left<{#1}\right>}

\newcommand{\nnn}{\nonumber\\}
\usepackage{bm}
\usepackage{cite}
\usepackage{comment}

\begin{document}

\title{\bf Spontaneous CP violation, sterile  neutrino dark matter and leptogenesis}

\author{Yanjin Jiang\thanks{12345056@zju.edu.cn}}
\author{Norimi Yokozaki\thanks{n.yokozaki@gmail.com}}

\affil{{\small Zhejiang Institute of Modern Physics and Department of Physics, Zhejiang University, Hangzhou, Zhejiang 310027, China}}

\date{}

\maketitle

\begin{abstract}
\noindent
We constructed a model for spontaneous CP symmetry breaking in five-dimensional space-time that has a potential to solve the strong CP problem. To explain the nature of dark matter and the baryon asymmetry of the universe, three right-handed neutrinos and $U(1)_{B-L}$ gauge interaction are introduced in the bulk, in addition to the field contents of the Bento-Branco-Parada model. The wave-function profiles in the fifth dimension can suppress dangerous operators allowed by symmetries, and the scale of spontaneous CP symmetry breaking can be sufficiently large to be consistent with thermal leptogenesis.
In this model, the lightest right-handed neutrino serves as dark matter with a mass of $\mathcal{O}(10)$ keV. This small mass and the necessarily small mixing are explained by the exponentially localized wave-function in the fifth dimension due to a bulk mass term. The correct relic abundance is achieved thanks to the $U(1)_{B-L}$ gauge interaction. The other heavy right-handed neutrinos explain the baryon asymmetry of the universe through leptogenesis, with the required CP violating phases generated by interactions involving heavy leptons.

\end{abstract}

\clearpage

\section{Introduction}

The Standard Model (SM) of particle physics is a highly successful theory that explains almost all observed phenomena up to $\mathcal{O}(100 \mathchar`- 1000)$ GeV. However, despite its success, the SM is not considered as the ultimate theory. There are several critical problems both phenomenologically and theoretically.

One major issue is that the SM does not include a candidate for dark matter, even though the existence of dark matter is considered robust and well-supported by observations. Additionally, in the SM, neutrinos are massless, whereas it is now well-established that neutrinos possess tiny but finite masses. 

The tiny neutrino masses are naturally explained by seesaw mechanism~\cite{seesaw_yanagida, Yanagida:1979gs, Gell-Mann:1979vob, Glashow:1979nm, Mohapatra:1979ia}, where the heavy right-handed neutrinos and their Majorana mass terms are introduced. It is also noted that the right-handed neutrinos with CP violating Yukawa couplings can explain the baryon asymmetry of the universe through leptogenesis~\cite{Fukugita:1986hr}.

Besides the phenomenological problems, there is a  significant theoretical issue in the SM, the strong CP problem. This problem arises because the strong CP-violating phase in the QCD Lagrangian must be unnaturally small, on the order of $\mathcal{O}(10^{-10})$ (see e.g., Refs.~\cite{Kim:2008hd,Hook:2018dlk} for review). If this phase were larger, it would induce a too larger electric dipole moment for the neutron, which has not been observed~\cite{Abel:2020pzs}.~\footnote{In the SM, the theta parameter is receive radiative correction at 7-loop level through CKM phase~\cite{Ellis:1978hq}.}

To address the strong CP problem, we explore the direction of spontaneous CP symmetry breaking. In this scenario, due to the exact CP symmetry of the Lagrangian, the strong phase is absent at the tree level, which provides a natural solution.

The Nelson-Barr mechanism offers a way to generate the CP violating phase in the CKM matrix, without reintroducing the too large strong CP violating phase~\cite{Nelson:1983zb,Barr:1984qx}. This approach effectively separates the origin of CP violation in the CKM matrix from the strong CP problem.

However, models based on spontaneous CP symmetry breaking have certain potential disadvantages compared to axion models (see e.g., Refs~\cite{Kim:2008hd,Hook:2018dlk} for review), where the global Peccei-Quinn symmetry is introduced~\cite{PhysRevLett.38.1440, PhysRevD.16.1791}.
In the axion models, the axion not only resolves the strong CP problem but also potentially serves as a dark matter candidate. Moreover, in the models of spontaneous CP symmetry breaking, generating the baryon asymmetry in the universe becomes more challenging as there is no CP violation at the Lagrangian level.

Additionally, dangerous operators that could generate an excessively large strong phase, force the scale of spontaneous CP symmetry breaking to be below $10^8$ GeV~\cite{Dine:2015jga}. This constraint gives a tension to models that explain baryon asymmetry through leptogenesis, as the masses of the right-handed neutrinos and the reheating temperature are required to be larger than about $10^9$\,GeV~\cite{Davidson:2002qv}. If this is the case, there is a cosmological domain wall problem due to the spontaneous break down of the discrete symmetry.  
To overcome these challenges, we have proposed a supersymmetric Nelson-Barr model~\cite{Evans:2020vil}. (See  Refs.~\cite{Cherchiglia:2020kut, Cherchiglia:2021vhe, Valenti:2021rdu,  Girmohanta:2022giy, Asadi:2022vys, Bai:2022nat, Pramanick:2022put, Nakagawa:2024ddd, Murai:2024alz} for recent studies of Nelson-Barr type models.) In this model, the scale of spontaneous CP breaking can exceed $10^{9}$ GeV, aligning with the requirements of thermal leptogenesis. The masses of the right-handed neutrinos are generated through their couplings with the CP symmetry breaking scalar, which also introduces the necessary CP violating phase for leptogenesis.

In this paper, we further develop this idea by proposing a model for spontaneous CP symmetry breaking within a five-dimensional space-time framework. This approach naturally accommodates right-handed neutrino dark matter with a mass of $\mathcal{O}(10)$ keV. The suppression of the mass, compared to those of other right-handed neutrinos, as well as the suppression of dangerous operators, is straightforward in this scenario, as the wave-function profile in the fifth dimension offers an alternative way of suppressing the operators for zero modes, beyond relying solely on symmetries.
   
\section{A model in five-dimensional space-time}

We extend the Bento-Branco-Parada model~\cite{Bento:1991ez} by including the right-handed neutrinos, $N_i\ (i=1 \dots 3)$ with $U(1)_{B-L}$ gauge symmetry. We introduce heavy leptons, $\eta$ and $\eta'$, in addition to heavy $SU(2)$ singlet quarks, $D$ and $D'$, that generates the CKM phase. 
The heavy leptons generate CP violating phases, which are required to explain the baryon asymmetry of the universe through leptogenesis~\cite{Fukugita:1986hr}.

To suppress dangerous operators that re-generates the strong CP phase larger than $\mathcal{O}(10^{-10})$, we consider a flat fifth dimension:~\footnote{We do not consider warped extra-dimension to avoid potential problems due to the lower-cutoff scale and lighter KK modes.} the SM Higgs and a scalar for spontaneous CP symmetry breaking are placed on different branes. This allows us to suppress the dangerous operators that are not forbidden by a typically introduced discrete symmetry (see appendix \ref{sec:dops}).

One of the right-handed neutrinos is a dark matter candidate with a $\mathcal{O}(10)$ keV mass that is naturally explained with an exponentially localized wave-function of the zero-mode in the extra-dimension~\cite{Kusenko:2010ik}. The action is given by
\begin{eqnarray}
	S = \int d^4 x \int_0^L d y  (\mathcal{L}_1 \delta(y) + \mathcal{L}_2 \delta(y-L)  + \mathcal{L}_{\rm bulk} ).
\end{eqnarray}
The SM Higgs is localized at the $y=0$ brane and a singlet scalar $S$ is localized at the $y=L$ brane; $S$ is responsible for spontaneous CP violation. The location of the fields is summarized in Table.~\ref{t:locations}. We may take $L^{-1} = 10^{15}\mathchar`- 10^{16}$\,GeV, and the five-dimensional Planck mass is $M_5 = (M_P^2 L^{-1})^{1/3} \approx (1.8\,\mathchar`-3.9) \times 10^{17}$ GeV.~\footnote{The field-theoretic cutoff scale is higher than \( M_5 \) because there are no strong coupling constants.}

The boundary conditions and $Z_2$ charge assignment are summarized in Table.~\ref{t:bcs} and \ref{t:charge}. We denote the boundary conditions as
\begin{eqnarray}
	&&	(+,+):\ \  \Psi_R|_{\rm boundaries}=0, \ \  (\partial_5 - M) \Psi_L|_{\rm boundaries}=0, \nnn  
	&&	(-,-):\ \  \Psi_L|_{\rm boundaries}=0, \ \ (\partial_5 + M) \Psi_R|_{\rm boundaries}=0,    
\end{eqnarray}
where $M$ is a bulk mass, and if $M \neq 0$, the zero-mode has an exponential profile with respect to the fifth dimension, $y$. Our model is identical to $S_1/Z'_2$ compactification by identifying $L=\pi R$, and $+$ and $-$ as $Z'_2$ parity.

As in the Bento-Branco-Parada model, $Z_2$ charge is introduced to distinguish the heavy quarks/leptons and $S$ from the SM quarks/leptons and Higgs fields. This \( Z_2 \) symmetry is anomaly-free and is considered a discrete gauge symmetry~\cite{Krauss:1988zc,Preskill:1990bm,Banks:1989ag}.

Interactions at the $y=0$ brane are given by 
\begin{eqnarray}
	-\mathcal{L}_{1} &\ni&  V(H) + V(\phi)+ y_{d,ij} \overline{Q_i} H d_{j,R} + h.c.\nonumber \\
	&+&
	y_{u,ij} \overline{Q_i} \tilde{H} u_{j,R} + h.c.\nonumber \\
	&+&
	y_{e,ij} \overline{L_i} H e_{j,R}+ h.c. \nonumber \\
	&+&
	y_{\nu,ij} \overline{L_i} \tilde{H} N_{j,R}+ h.c.\nonumber \\
	&+& 
	\frac{1}{2} \kappa_i \delta_{ij} \phi \overline{(N_{i,R})^c} N_{j,R} + h.c. \nnn
	&+& (\mu_D \overline{D_L} D'_{R} + \mu_\eta \overline{\eta_L} \eta'_R)+ h.c., 
\end{eqnarray}
where $V(H)$ is the SM Higgs potential, $V(H)= \lambda (|H|^2-v^2)^2$, and $V(\phi) = V(H\to \phi, v \to v_{B-L})$. We define $M_{N,i} \equiv \kappa_i v_{B-L}$. We neglect the mixing between $H$ and $\phi$ for simplicity.

Crucially, $N_1$ and $D'$ have bulk mass terms
\begin{eqnarray}
	\mathcal{L}_{\rm bulk}	\ni + m_b \overline{N}_1 N_1 - m'_b \overline{D'} D'
\end{eqnarray}
with $(-,-)$ boundary condition. As a result, the zero mode of $N_{1,R}$ ($D'_R$) has an exponential profile of wave-function in $y$ direction, $\exp(m_b y)$ ($\exp(-m'_b y)$), localizing at the $y=L$ brane ($y=0$ brane). See appendix \ref{sec:5dexp}.
The exponentially localized profile of the zero mode of $N_{1,R}$ naturally leads to the small Majorana mass of $\mathcal{O}(10)$ keV for $m_b L = \mathcal{O}(10)$.
The zero mode couplings and mass are suppressed as
\begin{eqnarray}
	y_{\nu,i1}^0 \simeq \frac{y_{\nu,i1} }{L} e^{-m_b L},\ 
	M_{N,1}^0 \simeq \frac{M_{N,1}} {L} e^{-2 m_b L}. \label{eq:n_mass}
\end{eqnarray} 

The left-handed (right-handed) SM fermions have the $(+,+)$ ($(-,-)$) boundary condition. The gauge bosons have $(+,+)$ boundary condition (Neumann boundary condition) for $A_\mu^a$ components and $(-,-)$ (Dirichlet boundary condition) for $A_5^a$ components so that $A_\mu^a$ have zero modes.  

At $y=L$, the singlet scalar $S$ has a potential that spontaneously breaks CP symmetry and has Yukawa interactions. The Yukawa interactions generate the CP violating phases in the quark sector and neutrino sector. The Lagrangian is given by
\begin{eqnarray}
	-\mathcal{L}_{2} &\ni& V(S) + 
	(k_i S + k'_i S^*) \overline{D_L} d_{i,R} + h.c. \nnn
	&+& (l_j S + l'_j S^*)  \overline{\eta_L}  N_{j,R} + h.c.
\end{eqnarray}
where the potential for $S$ can be written as
\begin{eqnarray}
	V(S) = |S|^2 (a_1 + b_1 |S|^2) 
	+ (S^2 + S^{*\, 2}) (a_2 + b_2 |S|^2) + b_3(S^4 + S^{*\, 4}).
\end{eqnarray}

With the above potential, $S$ acquires a complex vacuum expectation value, $\vevx{S}$. We assume $\vevx{S} > M_{N,2}$ for thermal leptogenesis to work. The scale of $\vevx{S}$ can be $10^{11} \mathchar`- 10^{13}$ GeV. The CP violating couplings to the right-handed neutrinos are required for leptogenesis.\footnote{The CP violating couplings to the SM leptons do not generate the required CP violating phase. (See appendix~\ref{sec:models}.)}

Let us briefly review the mixing and CP violation in the down-type quarks, focusing on the zero-modes. The right-handed quarks, $D'_R$ and $d_{i,R}$, mix significantly while the left-handed quarks, $d_{i,L}$ and $D_L$, have suppressed mixing. Still, the unsuppressed CP violating phase of the CKM matrix is generated. By defining ${\bm F}_i \equiv k_i^0 \vevx{S} + {k'}_i^{0} \vevx{S}^*$, $\hat{m}_d = {\rm diag}(m_d, m_s, m_b)$ and $(m_d)_{ij} = y_{d,ij}^0 v$, the CKM matrix is expressed as
\begin{eqnarray}
	V_{\rm CKM} \hat{m}_d^2 V^{-1}_{\rm CKM}
	\simeq m_d m_d^\dag - 
	m_d \bm{F}^* \bm{F}^T m_d^\dag \, (\mu_D^2 + \bm{F}^T \bm{F}^*)^{-1} 
\end{eqnarray}
where
\begin{eqnarray}
 \left(
 \begin{array}{c}
 	d_{i,L}\\
 	D_L
 \end{array}
 \right)	= 
 \left(
 \begin{array}{cc}
 V_{CKM}	& s_1 \\
 	s_2 & s_3
 \end{array}
 \right)
 \left(
\begin{array}{c}
	\tilde{d}_{i,L}\\
	\tilde{D}_L
\end{array}
\right), \ \ s_3 \approx 1, \ s_{1,2} \approx 0 \ \ {\rm for} \vevx{S} \gg v\,.
\end{eqnarray}
Here, the fields with tilde are mass eigenstates.

The Lagrangian $\mathcal{L}_2$ may contain a dangerous operator $\rho_1 M_*^{-1}(S^2 + \kappa_2 S^{2\,*}) \overline{D}_L D'_R + h.c.$, where $M_* \sim M_5$ is a cutoff scale. For $m'_b L \sim 20$, $\rho_1 \sim e^{-m'_b L} \sim 10^{-9}$. Together with the suppression of $|\vevx{S}|/M_*$, this operator is safe.

\begin{table}
	\caption{Location of field contents.}
	\begin{center}
		\begin{tabular}{c|c|c}
			brane ($y=0$) & bulk & brane ($y=L$) \\
			\hline
			$H$, $\phi$, $\mu_{D,\eta}$ &
			$D$, $D'$, $\eta$, $\eta'$, $N_i$, SM matter fields and gauge fields & 
			$S$
		\end{tabular}
	\end{center}
	\label{t:locations}
\end{table}

\begin{table}
	\caption{Boundary conditions for left-handed components. The left-handed components of $D$ and $E$ and the right-handed components of $D'$, $E'$ and $N_1$ have zero modes while the others do not.}
	\begin{center}
		\begin{tabular}{c|cccccccccccccc}
			& $D$ & $D'$ & $\eta$ & $\eta'$ & $N_i$
			\\
			\hline \hline
			& $(+,+)$ & $(-,-)$ & $(+,+)$ & $(-,-)$ &  $(-,-)$ \\
		\end{tabular}
	\end{center}
	\label{t:bcs}
\end{table}

\begin{table}
		\caption{$Z_2$ charge.}
\begin{center}
\begin{tabular}{c|cccccccccccccc}
	& $D$ & $D'$ & $\eta$ & $\eta'$ & $S$ & the others
	 \\
	\hline \hline
	$Z_2$ & 1 & 1 & 1 & 1 & 1 & 0 \\
\end{tabular}
\end{center}
\label{t:charge}
\end{table}

\subsection{CP violating phases in the neutrino sector}

In order to make $N_1$ as a sterile neutrino dark matter, $l_1$ and $l'_1$ are assumed to be negligibly small. 
This is justified when $l_i$ and $l'_i$ are condensation of scalar fields, and $l_1$ and $l'_1$ are (exponentially) localized at $y=0$ brane. (For a setup, see e.g. Ref.~\cite{Flacke:2008ne,Haba:2009uu} and appendix~\ref{sec:5dexp2}.) 
In this case, we may extend $Z_2$ symmetry to $Z_4$ symmetry, and assign  
\begin{eqnarray}
	&& Z_4(\eta)=Z_4(\eta')=1, \ 
	Z_4(l_i)=Z_4(l'_i)=-1, 
	\nnn	&&  Z_4(S)=Z_4(D)=Z_4(D')=2.
\end{eqnarray}

After the symmetry breaking of $U(1)_{B-L}$, 
\begin{eqnarray}
-\mathcal{L} &\ni&	\overline{\eta_L} (C_2 N_{2,R} + C_3  N_{3,R} + \mu_\eta \eta'_R) + h.c.
\nnn	&+& 	\sum_{i=2,3} \frac{1}{2}M_{N,i} \overline{(N_{i,R})^c} N_{i,R} + h.c.,
\end{eqnarray}
where $C_i \equiv l_i^0 \vevx{S} + {l'}_i^0 \vevx{S}^*$, and 
$C_2 \sim C_3 \sim \mu_\eta \gg M_{N,2}, M_{N,3}$ is assumed.
The mass term can be written as
\begin{eqnarray}
	\frac{1}{2} (\overline{(N_{2,R})^c}\ \overline{(N_{3,R})^c}\ \overline{({\eta'_R})^c}\ \overline{\eta_L})
	M_{N\eta}
	\left(
	\begin{array}{c}
		N_{2,R}\\
		N_{3,R}\\
		\eta'_R \\
		(\eta_L)^c
	\end{array}
	\right) + h.c.,
\end{eqnarray}
where
\begin{eqnarray}
	M_{N\eta} = 
	\left(
	\begin{array}{cccc}
	 M_{N,2}	& 0 & 0 & C_2\\
		0 & M_{N,3} & 0 & C_3 \\
	0 & 0 & 0 & \mu_\eta  \\
	C_2 & C_3 & \mu_\eta & 0 
	\end{array}
	\right). \label{eq:mn4}
\end{eqnarray}
Since ${\rm Det}(M_{N\eta}^\dag M_{N\eta}) = M_{N,2}^2 M_{N,3}^2\mu_{\eta}^4$, there are two lighter states of the order of $M_{N,2}$ and $M_{N,3}$, and the other two states have masses of the order of $\mu_\eta$.

\begin{figure}
	\begin{center}
		\includegraphics[width=0.8\textwidth]{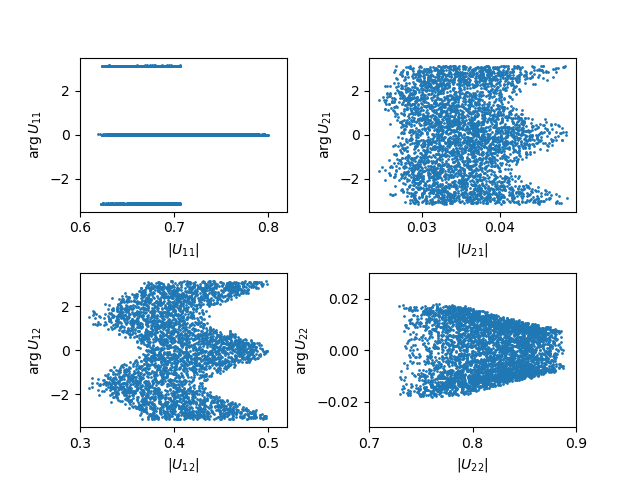}		
	\end{center}
	\caption{The elements of the unitary matrix.}
	\label{fig:umat}
\end{figure}

Using the unitary matrix, the mass matrix in Eq.~\eqref{eq:mn4} is diagonalized:
$U^T M_{N\eta} U = {\rm diag}(M_2, M_3, M_4, M_5 )$, $M_{4,5} \sim \mu_\eta \gg M_{2,3}$.
The 1,2 elements of the unitary matrix are important for the parameter region of our interest. We define
$u_{\alpha \beta} \equiv U_{\alpha \beta} \ (\alpha,\beta=1,2)$, and $u$ satisfies 
\begin{eqnarray}
	u \hat{M}_{N}^2 u^{-1} =
	(z y^{-1} + y^{-1} x )^{-1} (z y^{-1} x + y^\dag),
\end{eqnarray}
where
\begin{eqnarray}
&&	x = \left(
	\begin{array}{cc}
	M_{N,2}^2+|C_2|^2	&  C_2^* C_3\\
	C_2 C_3^*	& M_{N,3}^2+|C_3|^2
	\end{array}
	\right), \ \ 
	y =
	\left(
	\begin{array}{cc}
	C_2^* \mu_\eta	&  C_2 M_{N,2}\\
	C_3^* \mu_\eta	&  C_3 M_{N,3}
	\end{array}
	\right), \ \ \nnn
&&	z=
	\left(
	\begin{array}{cc}
	\mu_\eta^2	& \\
		& \mu_\eta^2 + |C_2|^2 + |C_3|^2
	\end{array}
	\right), \ \ 
	\hat{M}_N^2 =
	\left(
	\begin{array}{cc}
	M_{2}^2 &	\\
		& M_{3}^2
	\end{array}
	\right).
\end{eqnarray}
Because of $C_2$ and $C_3$, 
$u$ has complex phases. In fig.~\ref{fig:umat}, the absolute value and the phases of the elements of $u$ are shown. We fix $M_{N,2}/\mu_{\eta}=10^{-2}$ and $M_{N,3}/\mu_{\eta}=10^{-1}$, and take $C_2$ and $C_3$ randomly in the following ranges: 
\begin{eqnarray}
	&&|C_2/\mu_\eta| = [0.8, 1.2],\  |C_3/\mu_\eta| = [0.8,1.2], \nnn 
	&&{\rm arg}(C_2)= [-\pi, \pi], \ {\rm arg}(C_3)= [-\pi, 
	\pi].
\end{eqnarray}
It can be seen that $U_{21}$ and $U_{21}$ have $\mathcal{O}(1)$ phases, and $U_{22}$ has $\mathcal{O}(10^{-2})$ phase. Here, $U_{11}$ also has $\mathcal{O}(10^{-2})$ phase. With these phases, the baryon asymmetry of the universe can be explained through leptogenesis~\cite{Fukugita:1986hr}. 

The Yukawa couplings in the mass eigen basis are 
\begin{eqnarray}
	\tilde{y}_{\nu,i1} &=& {y}_{\nu,i1} \nnn
	\tilde{y}_{\nu,i2} &=& {y}_{\nu,i2} U_{11} + {y}_{\nu,i3} U_{21} \nnn 
	\tilde{y}_{\nu,i3} &=& {y}_{\nu,i2} U_{12} + {y}_{\nu,i3} U_{22} \nnn
	\tilde{y}_{\nu,i4} &=& {y}_{\nu,i2} U_{13} + {y}_{\nu,i3} U_{23} \nnn
	\tilde{y}_{\nu,i5} &=& {y}_{\nu,i2} U_{14} + {y}_{\nu,i3} U_{24}\, . \label{eq:n_yukawa} 
\end{eqnarray}
Hereafter, we omit the index 0 indicating the zero mode; unless stated otherwise, couplings refer to zero modes.
By taking $y_{\nu,i1}=0$, the rank of $\tilde{y}_{\nu}$ is 2, and there are two massive left-handed neutrinos. 
The tiny $y_{\nu,i1}$ is required to satisfy gamma ray constraints when $N_1$ is considered to be dark matter.
Consequently, one of the left-handed neutrinos is massless.
For $M_{4,5} \gg M_{2,3}$, the contributions involving $\tilde{y}_{\nu,i4}$ and $\tilde{y}_{\nu,i5}$ are suppressed and negligible.

\section{Sterile neutrino dark matter and leptogenesis}

\subsection{Sterile neutrino dark matter}
In our model, the lightest right-handed neutrino, $N_1$, has exponentially suppressed mass compared to $N_2$ and $N_3$ as shown in Eq.~\eqref{eq:n_mass}. The right-handed neutrino, or sterile neutrino, with a mass of $\mathcal{O}(10)$ keV is a dark matter candidate. 
The sterile neutrino mixes with the left-handed neutrinos, and the size of the mixing is
\begin{eqnarray}
	\theta^2 &\approx& \frac{\sum_i |\tilde{y}_{\nu,i1}|^2\vevx{H^0}^2}{M_{1}^2} 
	\\&\approx& 7.6 \times 10^{-14}
	\left(\frac{\sum_i |y_{\nu,i1}|^2}{10^{-27}}\right)
	\left(\frac{20\,\rm keV}{M_{1}}\right)^2, \ \ M_1 \equiv M_{N,1}.
\end{eqnarray}
From gamma ray constraints ($\nu_s \to \nu_{e,\mu,\tau} + \gamma$), the mixing angle needs to be smaller than 
$\sin^2(2\theta) \lesssim 10^{-12}\  (10^{-13})$ for 10\,keV (20\,keV) ~\cite{Roach:2022lgo, Zakharov:2023mnp}.
With such small mixing, it is difficult to produce right amount of the sterile neutrino dark matter through the mixing to left-handed neutrinos~\cite{Dodelson:1993je}. The required Yukawa couplings to avoid the gamma-ray constraints are tiny. Therefore, the one of the light neutrinos is essentially massless. 

\vspace{12pt}

Now, we focus on production of $N_1$ and $N_2$  from $U(1)_{B-L}$ gauge interaction:
$l \bar{l}, q\bar{q} \to N_{1,2} \bar{N}_{1,2}$. Here, $v_{B-L}, M_{3} > T_R> M_{1}, M_{2}$ is assumed, and $N_3$ are not produced in the early universe. (The assumption of $M_3 > T_R$ is just for simplicity, and is not essential.)
Here, $T_R$ is the reheating temperature after inflation.
The discussion about cosmological evolution has some similarity with Ref.~\cite{Dunsky:2020dhn}.

\paragraph{Thermalied case}

By comparing the interaction rate of $l \bar{l}, q\bar{q} \leftrightarrow N_{1,2} \bar{N}_{1,2}$, $\Gamma_{int} \sim T^5/v_{B-L}^4$, and the Hubble rate in the radiation-dominated era, 
\begin{eqnarray}
	\ H = \sqrt{\frac{\pi^2 g_* (T)}{90}} \frac{T^2}{M_P},
\end{eqnarray}
the condition for the thermalization of $N_1$ and $N_2$ is obtained: $N_1$ and $N_2$ are in thermal bath when the temperature is larger than  
\begin{eqnarray}
	T_{th} \equiv  10^{12}\,{\rm GeV} \times \left( \frac{v_{B-L}}{10^{13} \rm GeV} \right)^{4/3}.
\end{eqnarray}
For $T > T_{th}$, 
the number density over entropy density ($n/s$) of $N_1$ and $N_2$, are fixed as $Y_{1} = Y_{2} = Y \approx 0.004$; $Y$ remains constant until $N_2$ decays. If there is no dilution, the energy density of $N_1$ is too large to explain the observed dark matter abundance unless $M_{1}$ is inconsistently small as $M_{1} = \mathcal{O}(100)$\,eV~\cite{PhysRevD.96.023522}.  

When the universe becomes colder, $N_2$ dominate the energy density of the universe as the radiation energy decrease faster than the matter energy density.
The energy density of $N_2$ dominates when the temperature becomes smaller than
\begin{eqnarray}
	T_{dom} = \frac{4}{3} M_{2} Y_{2}.
\end{eqnarray}
It subsequently decays with the decay rate
\begin{eqnarray}
	\Gamma_2 \approx \sum_i |\tilde{y}_{\nu,i2}|^2 M_{2}/(4\pi),
\end{eqnarray}  
and produces entropy. As a result the abundance of $N_1$ (and the baryon asymmetry) is diluted as
\begin{eqnarray}
	\frac{\rho_{N_1}}{s} = M_{1} Y \frac{s_{\rm before}}{s} 
	\simeq \frac{3}{4} M_{1} \frac{T_{N_2}}{M_{2}} ,
\end{eqnarray} 
where
\begin{eqnarray}
T_{N_2} = \left(\frac{10}{\pi^2 g_*}\right)^{1/4} \sqrt{\Gamma_2 M_P}.	
\end{eqnarray}
We may require $T_{dom} > T_{N_2}$.
The abundance of $N_1$ over the critical density is
\begin{eqnarray}
	\Omega_{DM} = \frac{\rho_{N_1}}{\rho_{cr,0}}
	\approx 0.28 \times  \cdot \left( \frac{M_{1}}{15 \rm keV} \right) \left(\frac{10^{12} \rm GeV}{M_{2}}\right)^{1/2} 
	\left(\frac{\sum |\tilde{y}_{\nu,i2}|^2}{10^{-13}}\right)^{1/2} . \label{eq:omega1}
\end{eqnarray}
Although the observed dark matter energy density~\cite{ParticleDataGroup:2024cfk} can be explained, the Yukawa couplings are too small to explain the observed neutrino masses.~\footnote{The rank of $\tilde{y}_\nu$ is essentially one, and the only one massive neutrino exits in this case.} Therefore, we do not further investigate this possibility.

\paragraph{Non-thermalized case}

When the reheating temperature after the inflation is smaller than $T_{th}$, $N_1$ is not thermalized but they are produced from the SM particles in the thermal bath. The production is most efficient for the highest temparature.  The number density is estimated as~\cite{Khalil:2008kp}
\begin{eqnarray}
	Y \approx 8.8 \times 10^{-6} \left(\frac{T_R}{10^{10} {\rm GeV}}\right)^3 \left( \frac{3\times 10^{12}\,{\rm GeV}}{v_{B-L}}\right)^4.
\end{eqnarray}
The energy density of $N_1$ normalized by the critical density is 
\begin{eqnarray}
	\Omega_{N_1} \approx 0.26 \times 
	\left( \frac{M_{1}}{50\, {\rm  keV}} \right)
	\left(\frac{T_R}{10^{10} {\rm GeV}}\right)^3 \left( \frac{3\times 10^{12}\,{\rm GeV}}{v_{B-L}}\right)^4 \,. \label{eq:dm_relic}
\end{eqnarray}
The observed relic abundance of the dark matter is explained. Since the Yukawa couplings, $\tilde{y}_{\nu,i2}$, is not required to be tiny, the neutrino masses can be explained and leptogenesis can explain the observed baryon asymmetry.  

\subsection{Leptogenesis}
With the Yukawa couplings in Eq.~\eqref{eq:n_yukawa}, the  decays of $N_2$ can produce the lepton asymmetry. %The loop effect involving $N_3$ is also important as the interference between tree level and one-loop diagrams is required. 
The primordial lepton asymmetry is transferred into the baryon asymmetry through the sphaleron transitions. The baryon asymmetry is given by~\cite{Giudice:2003jh} 
\begin{eqnarray}
	Y_B \simeq - \frac{28}{79}Y_L, \ \ Y_L = \epsilon \,\eta\, Y_2^{\rm eq},
\end{eqnarray}
where $\eta$ is an efficient factor, which factorizes the number density and out-of-equilibrium decay. The parameter, $\epsilon$ represents a lepton asymmetry from the decay:
\begin{eqnarray}
	\epsilon \equiv \frac{\Gamma(N_2 \to L H)-\Gamma(N_2 \to \overline{L} \overline{H})}{\Gamma(N_2 \to L H)+\Gamma(N_2 \to \overline{L} \overline{H}))}
\end{eqnarray}
The observed value of $Y_B$ is $\approx 8 \times 10^{-11}$~\cite{Planck:2018vyg}.
Related to the washout effect, we define
\begin{eqnarray}
	\tilde{m} = \sum_i |\tilde{y}_{\nu,i2}|^2 v^2/M_{2}.
\end{eqnarray}
Here, $\tilde{m} = \mathcal{O}(10^{-2})$ eV is required to explain the neutrino masses.
The efficiency factor can be estimated as~\cite{Giudice:2003jh}
\begin{eqnarray}
	\frac{1}{\eta} = \frac{3.3 \times 10^{-3}\, {\rm eV}}{\tilde{m}} 
	+ \left( \frac{\tilde{m}}{0.55 \times 10^{-3}\, {\rm eV}}\right)^{1.16}.
\end{eqnarray}
For $\tilde{m} = 10^{-2}$ eV, $\eta \approx 0.03$ and $\eta Y_2 \approx 10^{-4}$. In this case, $\epsilon \approx -1.7 \times 10^{-6}$ explains the observed baryon asymmetry.

The asymmetry factor, $\epsilon$, is given by
\begin{eqnarray}
	\epsilon &=& \frac{1}{8\pi} \frac{\sum_{k=3,4,5}{\rm Im}[(\tilde{y}_\nu^T \tilde{y}_\nu^*)_{k2}^2 ]}{(\tilde{y}_\nu^T \tilde{y}_\nu^*)_{22}} g(x) \nnn
	&\simeq& -\frac{3}{16\pi}
	\frac{{\rm Im}[(\tilde{y}_\nu^T \tilde{y}_\nu^*)_{32}^2 ]}{(\tilde{y}_\nu^T \tilde{y}_\nu^*)_{22}}
	\frac{M_2}{M_3} \nnn 
    &=&  -\frac{3}{16\pi} 
		{\rm Im}[(\tilde{y}_\nu^T \tilde{y}_\nu^*)^2_{32} ] \frac{v^2}{\tilde{m} M_3} ,
\end{eqnarray}
where $x=M_{k}^2/M_{2}^2$, and $g(x) \to -\frac{3}{2\sqrt{x}}$ for $x \gg 1$. For the approximation of the second line, we use $M_{4,5} \gg M_{2,3}$ and $M_{3}^2/M_2^2 \gg 1$. By writing
\begin{eqnarray}
\frac{(\tilde{y}_\nu^T \tilde{y}_\nu^*)^2_{32}}{M_2 M_3} = k \cdot  \frac{\tilde{m} m_{\nu,3}}{v^4}  e^{i \theta},
\end{eqnarray}
the asymmetry is
\begin{eqnarray}
	\epsilon \simeq -k \cdot \frac{3}{16\pi} \frac{m_{\nu,3} M_2}{v^2} \sin\theta.
\end{eqnarray}
Here, $m_{\nu,3} \approx 0.05$\,eV~\cite{ParticleDataGroup:2024cfk} is the heaviest neutrino mass. For $k=\sin\theta=1$ and $M_2 = 2 \times 10^{10}$\,GeV, the observed baryon asymmetry $Y_B$ is explained. 
We may take $T_R = 4 \times 10^{10}$ GeV and $v_{B-L} = 8 \times 10^{12}$\,GeV, and then, the correct dark matter abundance is also explained for $M_1 \approx 40$\,keV (see Eq.\eqref{eq:dm_relic}).

\section{Conclusion}

We have constructed a model for spontaneous CP symmetry breaking in an extra-dimensional setup, including right-handed neutrinos and $U(1)_{B-L}$ gauge interaction. 
The proposed model not only solves the strong CP problem but explains the dark matter nature and the baryon asymmetry of the universe through  leptogeneis. By utilizing geometrical separation and localization of the fields, the dangerous operators that re-generates the too large strong phase are suppressed.
The lightest right-handed neutrino naturally has a mass of $\mathcal{O}(10)$\,keV as well as suppressed mixing with the SM neutrinos due to the exponentially localized wave-function in the fifth-dimension. 
The correct relic abundance is obtained through $U(1)_{B-L}$ gauge interaction. The required CP violating phases for leptogenesis are generated by the interactions involving heavy leptons. 

Although the mixing between the sterile neutrino dark matter and SM neutrinos is small enough to avoid current experimental bounds, the gamma-ray signal from its decay might be observed in the future.

\section*{Acknowledgments}
N. Y. is supported by a start-up grant from Zhejiang University. 
Y. J. is supported by Zhejiang University.

\appendix

\section{Dangerous operators} \label{sec:dops}

There are dangerous operators allowed by $Z_2$ symmetry:
\begin{eqnarray}
	&&\frac{S + S^*}{M_*} \overline{Q_i} H D_R',
	\ \ 
		\frac{S^2 + S^{2\,*}}{M_*^2} \overline{Q_i} H d_{j,R}, \ \ 
		\frac{S^2 + S^{2\,*}}{M_*^2} \overline{Q_i} \tilde{H} u_{j,R}, \ \ 	
				(\lambda_{sh,1} S^2 + \lambda_{sh,1} S^{*\,2} 
				+ \lambda_{sh,2} |S|^2) |H|^2,	
	\nnn &&	\frac{S^2 + S^{2\,*}}{M_*^2} \overline{D_L} D'_{R}, \ \ 
\end{eqnarray}
where the coefficients are abbreviated except for the fourth operator. 
The fourth operator generates the strong phase at the one-loop level through corrections to the down-type quark mass matrix. It also  contributes to the Higgs mass and electroweak symmetry breaking scale: the observed values are obtained through small couplings and/or cancellation with the tree-level mass parameters.
The one-loop corrections from zero-modes to these couplings, $\lambda_{sh,1}$ and $\lambda_{sh,1}$ are
\begin{eqnarray}
	\lambda_{sh,1} &\approx& 
	\frac{3}{8\pi^2} \left(\tr(Y_d \bm{k} \bm{k}^{'\,T} Y_d^T ) +
	\tr(Y_d \bm{k}' \bm{k}^{T} Y_d^T ) \right) \ln \frac{L^{-1}}{\mu_D}	\nnn
	&+&
	\frac{1}{8\pi^2} \left(\tr(Y_{\nu} \bm{l} \bm{l}^{'\,T} Y_\nu^T ) +
	\tr(Y_\nu \bm{l}' \bm{l}^{T} Y_\nu^T ) \right) \ln \frac{L^{-1}}{\mu_\eta},	\nnn
	\lambda_{sh,2} &\approx& \frac{3}{4\pi^2} \left(\tr(Y_d \bm{k} \bm{k}^{T} Y_d^T ) +
	\tr(Y_d \bm{k}' \bm{k}^{'\,T} Y_d^T ) \right) \ln \frac{L^{-1}}{\mu_D}	\nnn 
	&+&
	\frac{1}{4\pi^2} \left(\tr(Y_{\nu} \bm{l} \bm{l}^{T} Y_\nu^T ) +
	\tr(Y_\nu \bm{l}' \bm{l}^{'\,T} Y_\nu^T ) \right) \ln \frac{L^{-1}}{\mu_\eta}.
\end{eqnarray}
Therefore, the contribution that shifts the $\theta$-parameter is generated at two-loop level. It requires $k_3 \sim 0.1$.~\footnote{Alternatively, we may need a slight overlap of the wave function of zero modes for $S$ and $H$ to cancel the tree-level values and radiative corrections at some level. In that case, they are not fully localized on different branes.}

Except for the last operators, they are suppressed by localizing $H$ and $S$ on different branes. 
The last operator is suppressed if $D_R'$ is (exponentially) localized on the brane where $S$ does not reside.
\section{Localized bulk fermion}\label{sec:5dexp}

Here, we show that the zero-mode of a bulk fermion has exponential profile with respect to $y$, if there is a bulk mass.
Let us consider the free fermion in the bulk. 
\begin{eqnarray}
S = \int d^4 x \int_0^L d y \sqrt{g} \left[\frac{i}{2}(\overline{\Psi}  \gamma^A \partial_A  \Psi -  \partial_A \overline{\Psi}  \gamma^A   \Psi)-M \overline{\Psi} \Psi \right],
\end{eqnarray}
where $\Gamma^A = (\gamma^\mu, i \gamma_5)$ and $g^{MN}=(1,-1,-1,-1,-1)$. 
The equations of motions lead to
\begin{eqnarray}
&&	[i \gamma^\mu \partial_\mu P_L - (\partial_5 + M) P_R] \Psi =0
\nnn && [i \gamma^\mu \partial_\mu P_R + (\partial_5 - M) P_L] \Psi =0
\end{eqnarray}
The minimization condition, $\delta S =0$, is satisfied for 
\begin{eqnarray}
	&&	(+,+):\ \  \Psi_R|_{\rm boundaries}=0, \ \  (\partial_5 - M) \Psi_L|_{\rm boundaries}=0, \ \ {\rm or} \nnn  
	&&	(-,-):\ \  \Psi_L|_{\rm boundaries}=0, \ \ (\partial_5 + M) \Psi_R|_{\rm boundaries}=0.    
\end{eqnarray}

KK decomposition for fermions
\begin{eqnarray}
	\Psi_{R} &=& \frac{1}{\sqrt{L}}\sum_n f_n(y) \psi^n_{R}(x),  
\nnn 
	\Psi_{L} &=& \frac{1}{\sqrt{L}}\sum_n g_n(y) \psi^n_{L}(x),  
\end{eqnarray}
where we require
\begin{eqnarray}
	(\partial_5 + M) f_n &=& m_n g_n, \nnn
	-(\partial_5 - M) g_n &=& m_n f_n, \nnn
	\frac{1}{L}\int dy f_m f_n &=& \delta_{mn}, \nnn
	\frac{1}{L}\int dy g_m g_n &=& \delta_{mn}, 
\end{eqnarray}
so that 
\begin{eqnarray}
	S_{\rm eff} = \int d^4 x \sum_n \left[\frac{i}{2}
	(\overline{\psi^n} \gamma^\mu \partial_\mu \psi^n - \partial_\mu \overline{\psi^n} \gamma^\mu \psi^n) - m_n \overline{\psi^n} \psi^n \right]. 
\end{eqnarray}
%and
%\begin{eqnarray}
%	\frac{1}{L} \int_0^L dy f_{n} f_{m} = \delta_{m n}.
%\end{eqnarray}

For the zero mode, $m_{n=0} = 0$,
\begin{eqnarray}
	f_0(y) = \sqrt{\frac{M/|M|}{1-e^{-2 M L}}} e^{-M y}, \ \ 
	g_0(y) = \sqrt{\frac{-M/|M|}{1-e^{2 M L}}} e^{M y},
\end{eqnarray}
except for trivial solutions. For the $(-,-)$ boundary condition, $g_0(y)=0$.

For KK modes, $f_n$ and $g_n$ satisfy 2nd order equations:
\begin{eqnarray}
	(\partial_5^2 + \tilde{m}_n^2) f_n &=& 0, \nnn
(\partial_5^2 + \tilde{m}_n^2) g_n &=& 0, \ \ \tilde{m}_n^2 = m_n^2-M^2.
\end{eqnarray}
For $\tilde{m}_n^2 > 0$, there are  
solutions which are superposition of 
\begin{eqnarray}
	\sin(\tilde{m}_n y), \ \ \cos(\tilde{m}_n y),
\end{eqnarray}
and the mass of the KK mode satisfies $\tilde{m}_n = n \pi/L$.

The solutions with the $(-,-)$ boundary condition are
\begin{eqnarray}
	f_n(y) &=& \frac{\sqrt{2} M}{m_n} \left[\frac{-\tilde{m}_n}{M} \cos(\tilde{m}_n y)+\sin(\tilde{m}_n y) \right], \nnn
g_n(y) &=& \sqrt{2} \sin(\tilde{m}_n y).	
\end{eqnarray} 

\section{Localized bulk scalar}\label{sec:5dexp2}

We discuss $y$ dependence of the bulk scalar. The action is given by
\begin{eqnarray}
	S &=& \int d^4 x \int_0^L d y \sqrt{g} \left[
	\partial_M \Phi^\dag \partial^M \Phi 
	-V_b(|\Phi|) \right.
\nnn	&+& \left. 
	(r_1\partial_\mu \Phi^\dag \partial^\mu \Phi-V_1(|\Phi|)) \delta(y)
	+(r_2\partial_\mu \Phi^\dag \partial^\mu \Phi-V_2(|\Phi|)) \delta(y-L) 
	\right].
\end{eqnarray}
The minimization, 
$\delta S=0$ leads to the equations of motion, 
\begin{eqnarray}
	\partial_M \partial^M \Phi + 
	\frac{\partial V_b}{\partial \Phi^\dag}=0,
	\label{eq:eom_vev}
\end{eqnarray}
and boundary conditions
\begin{eqnarray}	
	\left[
	+\partial_5 \Phi -  r_1 \partial_\mu \partial^\mu \Phi - \frac{\partial V_1}{\partial \Phi^\dag} \right]_{y=0} = 0,
\nnn
	\left[
-\partial_5 \Phi -  r_2 \partial_\mu \partial^\mu \Phi - \frac{\partial V_2}{\partial \Phi^\dag} \right]_{y=L} = 0 .	
\end{eqnarray}
Let us focus on the zero-mode, satisfying $\Box \Phi = 0$.  
With $V_b=M^2 |\Phi|^2$, the general solution of Eq.\eqref{eq:eom_vev} is given by
\begin{eqnarray}
	\Phi(x,y) = (A e^{M y} + B e^{-M y}) \frac{\phi_0(x)}{\sqrt{L}}.
\end{eqnarray}
The coefficients are determined by the boundary conditions. For instance, if $r_2=V_2=0$, the one of the solutions is $\Phi(x,y) = N (e^{-2 ML } e^{M y} + e^{-M y}) \phi_0(x)/\sqrt{L}$, which is localized at $y=0$.

\section{Another model for CP violation in the lepton sector} \label{sec:models}

We consider the following mass terms for the leptons:
\begin{eqnarray}
	- \mathcal{L} &\ni&  \mu_E \overline{E_L} E'_R +  (l_j S + l'_j S^*) \overline{E_L} e_{j,R} + y_{e,ij} \overline{L_i} H e_{j,R} + h.c.
	\\
	&+& y_{\nu,ij} \overline{L_i} \tilde{H} N_{j,R} + \frac{1}{2} M_{N,i} \delta_{ij} \overline{(N_{i,R})^c} N_{j,R} + h.c.
\end{eqnarray}
The above Lagrangian leads to the mass terms for the charged leptons as
\begin{eqnarray}
	- \mathcal{L} \ni  \mu_E \overline{E_L} E'_R +  B_j \overline{E_L} e_{j,R} + m_{e,ij} \overline{e_{i,L}} e_{j,R} + h.c. \, , \label{eq:ap_mass1}
\end{eqnarray}
where $m_{e,ij}= y_{e,ij} \vevx{H^0}$, and $B_i = l_i \vevx{S}
+ 
l'_i \vevx{S}^*$, and the mass matrix takes the following form:
\begin{eqnarray}
	&& M_e = \left(
	\begin{array}{cc}
		m_{e,ij}	&  0\\
		\bm{B}^T	&  \mu_E 
	\end{array}
	\right).
\end{eqnarray}
The mass matrix is diagonalized by two unitary matrices: 
\begin{eqnarray}
	\left(
	\begin{array}{c}
		e_L \\
		E_L
	\end{array}
	\right)= V
	\left(
	\begin{array}{c}
		\tilde{e}_L \\
		\tilde{E}_L
	\end{array}
	\right),
	\ \ 
	\left(
	\begin{array}{c}
		e_R \\
		E'_R
	\end{array}
	\right)= U
	\left(
	\begin{array}{c}
		\tilde{e}_R \\
		\tilde{E}'_R
	\end{array}
	\right),	
\end{eqnarray}
where 
\begin{eqnarray}
	V^\dag M_e U = {\rm diag}(m_e, m_\mu, m_\tau, \tilde{M}_E)
	= {\rm diag}(m_f, \tilde{M}_E), \ \
	\tilde{M}_E \simeq \sqrt{\mu_E^2 + \bm{B}^T \bm{B}^*}.
\end{eqnarray}
Therefore, 
\begin{eqnarray}
	V^\dag M_e M_e^\dag V = U^\dag M_e^\dag M_e U= {\rm diag}(m_f^2, \tilde{M}_E^2). 
\end{eqnarray}
%
%The mixing matrix for the left-handed leptons is determined by diagonalizing $M_e M_e^\dag$ while that for the right-handed leptons is determined by diagonalizing $M_e^\dag M_e$.
%
Note that 
$E_R'$ and $e_{j,R}$ have significant mixing while $e_{L,j}$ and $E_L$ do not mix significantly. 
%Still, the approximate unitary matrix used for diagonalization, $e_L = V_e \tilde{e}_L$, has complex phases. 
Let us define the unitary matrix for the left-handed leptons as
\begin{eqnarray}
	V = \left(
	\begin{array}{cc}
		V_e & K \\
		S^T & T 
	\end{array}
	\right),  	
\end{eqnarray}
and it satisfies
\begin{eqnarray}
	M_e M_e^\dag V 
	= V {\rm diag}(m_f^2, \tilde{M}_E^2). 
\end{eqnarray}
The components of the unitary matrix are determined by the following four equations:
\begin{eqnarray}
	&& m_e m_e^\dag V_e + m_e \bm{B}^* S^T = V_e m_f^2
	\\
	&& m_e m_e^\dag K + m_e \bm{B}^* T = K \tilde{M}_E^2
	\\
	&& (\mu_E^2 + \bm{B}^T \bm{B}^*) S^T + \bm{B}^T m_e^\dag  V_e = S^T m_f^2
	\\
	&& (\mu_E^2 + \bm{B}^T \bm{B}^*) T + \bm{B}^T m_e^\dag K = T \tilde{M}_E^2
\end{eqnarray}
Using
\begin{eqnarray}
	S^T = - \bm{B}^T m_e^\dag V_e ( (\mu_E^2 + \bm{B}^T \bm{B}^*) I - m_f^2 )^{-1}, \label{eq:ap_s}
\end{eqnarray}
We obtain
\begin{eqnarray}
	V_e m_f^2 V_e^{-1} &=& m_e m_e^\dag - m_e \bm{B}^* \bm{B}^T m_e^\dag V_e ( (\mu_E^2 + \bm{B}^T \bm{B}^*) I - m_f^2 )^{-1} V_e^{-1}
	\nonumber \\
	&\simeq&  
	m_e m_e^\dag - m_e \bm{B}^* \bm{B}^T m_e^\dag (\mu_E^2 + \bm{B}^T \bm{B}^*)^{-1},
\end{eqnarray}
where the CP violating phases are not suppressed. 
The other components are 
\begin{eqnarray}
	K = \mathcal{O}(m_f/\mu_E), \ T \approx 1.  
\end{eqnarray}
Using $V_e$, the MNS matrix is given by
\begin{eqnarray}
	V_{MNS} = V_e^\dag O_{\nu,L}.
\end{eqnarray}  
The rotational matrix $O_{\nu,L}$ is determined to diagonalize the neutrino mass matrix
\begin{eqnarray}
	\frac{1}{2} m_{\nu,ij} \overline{\nu_i^c} P_L \nu_j + h.c. \to 
	\frac{1}{2} \overline{\nu^c} O_\nu^T m_{\nu} O_\nu P_L \nu + h.c.
	=
	\frac{1}{2} \overline{\nu^c} m_{\nu, \rm diag} P_L \nu  + h.c.,
\end{eqnarray}
where
\begin{eqnarray}
	m_{\nu,ij} = -y_{\nu,ik} M_{N,k}^{-1} y_{\nu,kj}^T \vevx{H^0}^2.
\end{eqnarray}

%For the CKM matrix, there is only one physical phase, however, because of the Majorana mass terms, we can not absorbed the phases by field redefinition, $\nu_{i,L} \to e^{i\theta_i} \nu_{i,L}, \ N_{i,R} \to e^{i\theta_i} N_{i,R}$. 
%Therefore, MNS matrix has 3 phases. 

Similarly, the mixing matrix for the right-handed leptons is expressed as
\begin{eqnarray}
	U = \left(
	\begin{array}{cc}
		U_e	& u_1 \\
		u_2	& u_3
	\end{array}
	\right).
\end{eqnarray}
From $M_e^\dag M_e U = U {\rm diag} (m_f^2, \tilde{M}_E^2)$,
\begin{eqnarray}
	\mu_E \bm{B}^T U_e + \mu_E^2 u_2 = u_2 m_f^2 \to
	u_2 \simeq -\bm{B}^T U_e/\mu_E. \label{eq:ap_u2}
\end{eqnarray}
The light lepton mass matrix in Eq.\eqref{eq:ap_mass1} is, therefore, 
\begin{eqnarray}
	\mu_E S^* u_2 + S^* \bm{B}^T U_e + V_e^\dag m_e U_e = m_f.	
\end{eqnarray}
Using Eqs.\eqref{eq:ap_s} and \eqref{eq:ap_u2}, $V_e^\dag m_e U_e \simeq m_f$ at the leading order.

Now we take the basis, $L = V_e \tilde{L}$, which keeps $SU(2)$, we approximately obtain 
\begin{eqnarray}
	- \mathcal{L} \ni   
	(m_{f,ii}/v) \delta_{ij} \overline{L_i} H P_R e_j
	+ \tilde{y}_{\nu,ij} \overline{L_i} \tilde{H} P_R N_j
	+ \frac{1}{2} M_{N,ii} \delta_{ij} \overline{N_i^c} P_R N_j
	+h.c.,
\end{eqnarray}
where 
\begin{eqnarray}
	\tilde{y}_{\nu,ij} = V_{e,ik}^\dag y_{\nu,kj}. 
\end{eqnarray}
Here, we dropped tildes. The SM Yukawa couplings for the leptons are diagonal while the Yukawa couplings, $\tilde{y}_{\nu,ij}$, have CP violating phases and are non-diagonal.

The parametrization used in Ref.~\cite{Casas:2001sr}
\begin{eqnarray}
	\tilde{y}_{\nu}^\dag &=& \, v^{-1} 
	{\rm diag}(\sqrt{M_{N,1}},\sqrt{M_{N,2}},\sqrt{M_{N,3}}) \  R \ D_\nu^{1/2} \,	
	V_{MNS}^\dag ,
\end{eqnarray}
where $R$ is a real orthogonal matrix (as $y_\nu$ is real), and 
\begin{eqnarray}
	D_\nu^{1/2} &=& 
	\left\{ 
	\begin{array}{c}
		{\rm diag}(\sqrt{m_{\nu,1}}, \sqrt{m_{\nu,2}}, \sqrt{m_{\nu,3}})\  \ ({\rm Normal \ order}) \\
		{\rm diag}(\sqrt{m_{\nu,3}}, \sqrt{m_{\nu,1}}, \sqrt{m_{\nu,2}}) \ \  ({\rm Inverted\  order})
	\end{array}
	\right.	.
\end{eqnarray}

Although, the model can successfully explain the MNS matrix, with the complex phases, the phases do not play any roles in leptogenesis as $(\tilde{y}^\dag \tilde{y})_{ij}\ (i\neq j)$ are real.

\providecommand{\href}[2]{#2}\begingroup\raggedright\endgroup

\end{document}